\documentclass[prd,preprint]{revtex4}

\newcommand{\ch}[1]{{\color{black} #1}}

\usepackage{url}
	\newcommand{\hl}[1]{{\color{black} #1}}
	\usepackage{amsmath}
	\usepackage{amsfonts}
  	\usepackage{amssymb}
  	\usepackage{float}
	\usepackage{makeidx}
	\usepackage{amsfonts}
	\usepackage[ansinew]{}
	\usepackage[usenames,dvipsnames]{pstricks}
	\usepackage{booktabs}
	\usepackage{multirow}
	\usepackage{epsfig}
	\usepackage{comment}

\usepackage{xcolor} 

\newcommand{\be}{\begin{equation}}
\newcommand{\ee}{\end{equation}}
\newcommand{\bea}{\begin{eqnarray}}
\newcommand{\eea}{\end{eqnarray}}

\makeindex

\begin{document}


\title{$H_0$ tension or $M$ overestimation?}

\author{Brayan Yamid Del Valle Mazo${}^{2,3}$}
\author{Antonio Enea Romano${}^{1,2,3}$}
\author{Maryi Alejandra Carvajal Quintero${}^{2,3}$}

\affiliation{%
${}^{1}$Theoretical Physics Department, CERN,CH-1211 Geneva 23, Switzerland\\
${}^{2}$ICRANet, Piazza della Repubblica 10,I–65122 Pescara, Italy\\
${}^{3}$Instituto de Fisica,Universidad de Antioquia,A.A.1226, Medellin, Colombia
}%

\date{\today}

\begin{abstract}

There is a strong discrepancy between the value of the Hubble parameter $H_0^P$ obtained from large scale observations such as the Planck mission, and the small scale value $H_0^R$, obtained from low redshift supernovae (SNe).
The value of the absolute magnitude $M^{Hom}$ used as prior in analyzing observational data is obtained from low-redshift SNe, assuming a homogeneous Universe, but the  distance of the anchors used to calibrate the SNe to obtain $M$ would be affected by a local inhomogeneity, making it inconsistent to test the Copernican principle using $M^{Hom}$, since $M$ estimation itself is affected by local inhomogeneities.

We perform an analysis of the  luminosity distance of low redshift SNe, using different values of $M$, $\{M^P,M^R\}$, corresponding to  different values of $H_0$, $\{H_0^P,H_0^R\}$, obtained from the model independent consistency relation between $H_0$ and $M$ which can be derived from the definition of the distance modulus. We find that the value of $M$ can strongly affect the evidence of a local inhomogeneity. We analyze  data from the Pantheon catalog, finding no significant statistical evidence of a local inhomogeneity using the parameters $\{M^R,H_0^R\}$, confirming previous studies, while with $\{M^P,H_0^P\}$ we find evidence of a small local void, which causes an overestimation of $M^R$ with respect to $M^P$. 

An inhomogeneous model with the parameters $\{M^P,H_0^P\}$ fits the data better than  a homogeneous model with $\{M^R,H_0^R\}$, resolving the apparent $H_0$ tension. 
Using $\{M^P,H_0^P\}$, we obtain evidence of a local inhomogeneity with a density contrast $-0.140 \pm 0.042 $, extending up to a redshift of $z_v =0.056 \pm 0.0002$, in  good agreement with recent results of galaxy catalogs analysis \cite{wong2021local}.
\end{abstract}

\pacs{Valid PACS appear here}
\maketitle

\section{Introduction}
\hl{There is a discrepancy between the  large scale estimations based on the cosmic microwave background (CMB) radiation \cite{Akrami:2018odb}, and the value obtained analyzing low redshift SNe \cite{Riess:2016jrr}.} The SNe analysis is based on the assumption that the Universe is well described by a spatially homogeneous solution of the Einstein's equations, but only an unbiased analysis can actually confirm the validity of this  assumption. \hl{In the past inhomogeneities were studied before the discovery of dark energy \cite{Moffat:1994ma}, and then as a possible alternative to dark energy \cite{Mustapha:1997fjz,Hunt:2008wp,Shafieloo_2010,Celerier:1999hp,Enqvist:2006cg,GarciaBellido:2008nz}, but large void models without dark energy were shown to be incompatible with multiple observations \cite{Moss_2011}. The study of the effects of  inhomogeneities in presence of dark energy  was then performed \cite{Romano:2011mx,Romano:2010nc}, showing how they could lead to a correction of the apparent value of the cosmological constant, or affect  \cite{Romano:2016utn,Fleury:2016fda} the Hubble diagram. 
In this work we fit data without any homogeneity assumption. 
}

Our analysis is confirming the existence of a local under-density  surrounding us  in different directions \cite{Keenan:2013mfa,Chiang:2017yrq,wong2021local}.  The probability of formation of such an inhomogeneity is low in the $\Lambda \rm CDM$ framework, but modified gravity \cite{Haslbauer:2020xaa} could alleviate this problem.

Many different approaches to the explanation of the $H_0$ tension have been proposed \cite{Vagnozzi:2019ezj,DiValentino:2019jae,Conley:2007ng}. 
We do not propose any modification of the standard cosmological model, but  perform an unbiased analysis of SNe luminosity distance data, including  the effects of local  inhomogeneities.
If inhomogeneities were absent we should obtain a confirmation in our analysis.

Number count observations allow to measure directly the baryonic matter density but there are some difficulties in deducing the total density field from number counts, due for example to selection effects. For this reason another  possible alternative approach is to  reconstruct the total matter density  distribution from its effects on the luminosity distance of  standard candles \cite{Chiang:2017yrq}, and standard sirens \cite{Abbott:2017xzu}. 
 The Doppler effect is the main low redshift effect of inhomogeneities on the luminosity distance of the sources of electromagnetic waves, such as standard candles, \cite{Bolejko:2012uj,Romano:2016utn}, due to the peculiar velocity of the sources and the observer, and a similarly also for the luminosity distance of GW sources \cite{Bertacca:2017vod}.

It has been shown that \cite{Romano:2016utn} in the low redshift perturbative regime the monopole of the effects on the luminosity distance is proportional to the volume average of the density contrast. For an under-density this  corresponds to a peculiar velocity field pointing towards the outer  denser region, implying a local increase of the Hubble parameter, which could  account for the apparent difference between its  large and small scale estimation \cite{Romano:2016utn}. 
 
We adopt an unbiased approach, based on not assuming  homogeneity, and use the data to determine if the local Universe is homogeneous or not. 
Most of the effects of local inhomogeneities could  be removed by applying a redshift correction (RC), but the RC cannot remove all their effects if the depth of the galaxy catalog used for computing the RC is less than the size of the inhomogeneities \cite{Romano:2016utn}.

\section{Fitting of observational data}
\ch{In this papers we present the results of analyzing data from the Pantheon catalog \cite{Scolnic_2018} in the CMB frame. We have also analyzed data in the heliocentric reference frame, finding negligible differences.

This data is often analyzed under the assumption that all the effects of inhomogeneities have been removed by applying RC, but as explained in \cite{Romano:2016utn}, the 2M++ catalog \cite{2011MNRAS.416.2840L} used to estimate the peculiar velocity to obtain the RC, is not deep enough to eliminate the effects of an inhomogeneity extending beyond its depth $z=0.067$. The  edge of the inhomogeneity we obtain in our analysis is in fact  around the depth of 2M++. The effects of the homogeneity extend beyond the edge, as shown in  in Fig.(\ref{deltaz}).}

The observed quantity for SNe is the apparent magnitude $m$, while the luminosity distance $D_L$ is a derived quantity, and is model dependent, in the sense that it depends implicitly on $M$, which is one of the parameters of the model.

From the definition of distance modulus $\mu=m-M$ we have
\bea
D_L(z)&=&10^{\frac{\mu}{5}+1}=10^{\frac{m-M}{5}+1} \,,\label{DMU}
\eea
showing that an assumption for $M$ has to be made in order to get $D_L^{obs}$ from $m^{obs}$ . 
These are the main advantages of using $m$ :
\begin{itemize}
    \item It is not necessary to obtain $D_L$ from $m$ and compute the propagated errors.
    \item For different values of $\{H_0,M\}$ the data set of $m$ is the same, so the results can be plotted all together, while for $D_L$ a different dataset of each different $M$ has to be obtained from $m$.
\end{itemize}
Using $m$ makes clear  the distinction between observed data and parameters of the model, while when fitting $D_L$ the parameter $M$ is affecting both the model and the data $D_L^{obs}$, making the analysis less transparent. While theoretical predictions are  made in terms of $D_L(z)$,  observational data analysis in models with  varying $\{H_0,M\}$ should be more conveniently be performed in terms of $m$.

The theoretical model for the apparent magnitude $m^{th}$ is obtained from the theoretical luminosity distance $D_L^{th}$ 
\be
m^{th}=5 \log{D_L^{th}}-5 + M \,,
\ee
and the monopole effects of a local inhomogeneity are computed  using the formula \cite{Romano:2016utn}
\be
D_L^{th}(z)=\overline{D_L^{th}}(z)\left[1+\frac{1}{3} f\overline{\delta^{th}}(z) \right] \,, \label{Dlowz}
\ee
where $\overline{\delta}(z)$ is the volume averaged density contrast, $f$ is the growth factor, and $\overline{D_L^{th}}(z)$ is the luminosity distance of the background $\Lambda\rm CDM$ model. 
\hl{
 In  \cite{Romano:2016utn} (see Fig.(2) and section 6 therein)  it was shown that the above equations are in very good agreement with exact numerical calculations for the type of inhomogeneities studied in this paper.}

The different parameters are shown explicitly in these  equations
\bea
D_L(\Omega_i,H_0,I_i)&=&\overline{D_L}(\Omega_i,H_0)\left[1+\frac{1}{3} f\overline{\delta}(I_i) \right] \,,\\
m(\Omega_i,H_0,M,I_i)&=&5 \log{D_L(\Omega_i,H_0,I_i)}-5 + M \,,
\eea
where $I_i$ are the parameters modeling the inhomogeneity. A homogeneous model corresponds to $I_i=0$. Since no assumption about the homogeneity of the local Universe is made, if the Universe were homogeneous our analysis should confirm it.

As shown in \cite{Romano:2016utn,Chiang:2017yrq}, a  local inhomogeneity should only affect the luminosity distance locally, because far from the inhomogeneity the volume averaged density contrast of a finite size homogeneity is negligible, unless some higher order effect becomes dominant.

\section{The model of the local inhomogeneity}
We model the spherically symmetric local inhomogeneity with a density profile of the type
\be
\delta^{th}(\chi)=\delta_v [1-\theta(\chi-\chi_{v})] \,, \label{deltastep}
\ee
where $\delta_v$ is the density contrast inside the inhomogeneity, $\chi_v$ is the comoving distance of the edge of the inhomogeneity,  and $\theta(x)$ is the Heaviside function.
The volume averaged density contrast corresponding to the above profile is
\begin{equation}
\overline{\delta^{th}}(z)=
     \left\{
     \begin{array}{lcc}
     \delta_{v} &  & z<z_{v}  \, \\ 
     \delta_{v} \left[ \frac{z_v(1+z_v)}{z(1+z)} \right]^3   && z>z_{v} \,
     \end{array}
     \right \} \,,
\label{eq: delta ave piecewise}
\end{equation}
where $z_{v}$ is the inhomogeneity edge redshift. Details of the derivation of this formula are given in  appendix C. The above formulae are in agreement with the general result obtained in \cite{Romano:2016utn}, confirmed by numerical calculations for this kind of inhomogeneities, that the effects of low redshift inhomogeneities are suppressed at high redshift by the volume in the denominator of the volume average.

We minimize with respect to the two parameters $\delta_v,z_v$ the following $\chi^2(\delta_v,z_v)$

\begin{equation}
\chi^2 =\sum_{i,j} [m_i-m^{th}(z_i)]  C^{-1}_{ij} [m_j-m^{th}(z_j)]
\label{eq: chi2: m}
\end{equation}
where $C$ is the covariance matrix, $m_i$ and $z_i$ are the observed values of the apparent magnitude and redshift, and the sum is over all available observations.

\begin{figure}[H]
\centering
 \includegraphics[scale=0.37]{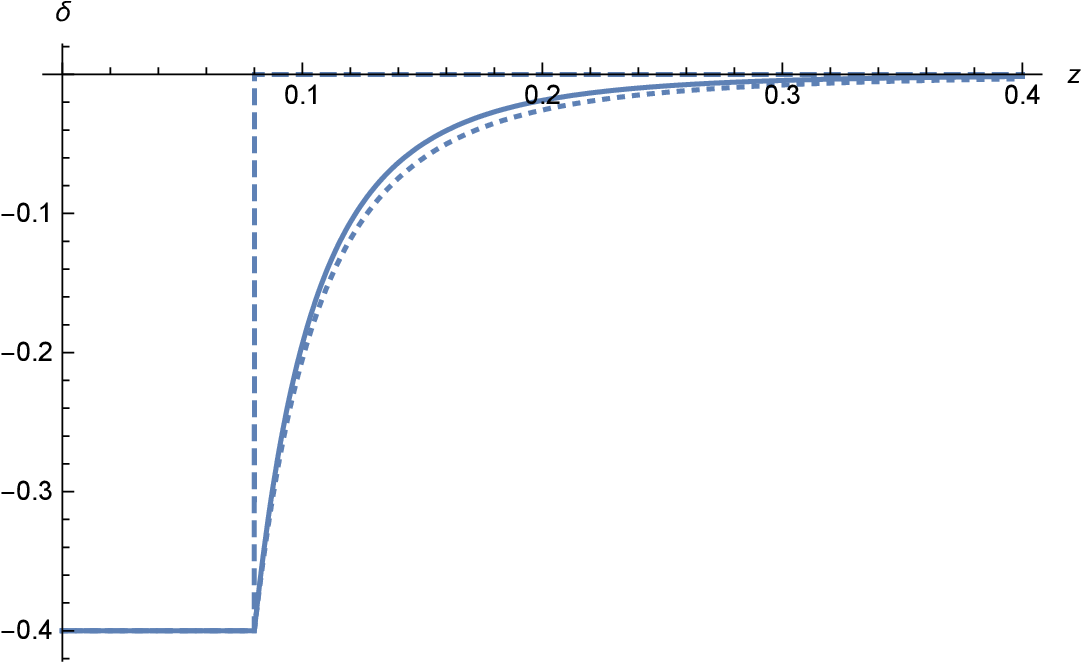}
 \caption[]{\ch{Plots of the step density contrast $\delta$ (dashed) defined in Eq.(\ref{deltastep}), its volume average $\overline{\delta}$ (solid)  in Eq.(\ref{eq: delta ave piecewise}), and the approximation (dotted) obtained by dropping the term $(1+z_v)/(1+z)$,  for $z_v=0.08$ and $\delta_v=-0.4$. The inhomogeneity effects are proportional to $\overline{\delta}$, and  extend beyond the edge of the void, but are suppressed beyond the edge, implying that high redshift observations are not affected by the local inhomogeneity.}}
 \label{deltaz}
\end{figure}

When using the  $H_0^R$ and $M^R$ our analysis gives results in agreement with previous studies such as \cite{Kenworthy:2019qwq}, not finding evidence of a local inhomogeneity, while when using $H_0^P$ and $M^P$ we obtain evidence of a local under-density remarkably similar to that found analyzing galaxy catalogs \cite{wong2021local}. 
The factor $(1+z)$ at low redshift can be safely neglected as shown in Fig.(\ref{deltaz}). 

\section{Fitting data assuming different values of $H_0$ and $M$}
The importance of the absolute magnitude in the analysis of SNe data was previously noted in \cite{Mazo:2019pzn,Efstathiou:2021ocp,Benevento:2020fev,Camarena:2021jlr}, and it plays an important role in assessing the presence of an inhomogeneity.
The approach adopted in this paper, consisting in using different priors for $\{H_0,M\}$  based on the model independent considerations given below, is only a first approximation. A full Bayesian analysis would be required to confirm the results, and we leave this to a future upcoming work.

From  the definition of distance modulus and Eq.(\ref{DMU}) we can derive a general model independent relation between different values of $\{H_0,M\}$ estimated at low redshift
\bea
M_{a} &=& M_{b} + 5\log_{10}\left(\frac{H_{a}}{H_{b}}\right) \,.
\eea
Details of the derivation are given in Appendix A.
For example this relation can be used  to obtain the implied Planck value $M^P$ from $H_0^P$ and $\{H_0^R,M^R\}$, which are the values obtained in \cite{Riess:2016jrr}. This is the value of $M$ which should be used when testing models with different values of $H_0$. Using $M^{R}$ as a prior when testing inhomogeneity, as done for example in \cite{Kenworthy:2019qwq,Camarena:2021mjr,Castello:2021uad}, is inconsistent, i.e. no local prior based on assuming homogeneity should be used when testing inhomogeneity. 

Another useful relation derived in the appendix is the one giving the correction to the absolute magnitude due to a local inhomogeneity
\bea
\Delta M &=& 5\log_{10} \left(1-\frac{1}{3}f \overline{\delta}(z)\right) \, \label{DM},
\eea
which shows how the absolute magnitude can be miss-estimated due to the unaccounted effects of a local inhomogeneity.
This type of relation for $M$ was used in \cite{Chiang:2017yrq} and more recently in \cite{Benevento:2020fev}. 
Using the above relations and the values obtained in \cite{Riess:2016jrr} as reference, we obtain $M^P$ from $H_0^P$, and fit $m^{obs}$ with different homogeneous and inhomogeneous models assuming  different values for $\{H_0,M\}$

In our notation $m^{\textup{Hom}}(H_0^R)$ and $m^{\textup{Inh}}(H_0^P)$ denote respectively a homogeneous model with $\{H_0,M\}=\{H_0^R,M^R\}$ and an inhomogeneous model with $\{H_0,M\}=\{H_0^P,M^P\}$.
We use the cosmological parameters $\Omega_i$ from the best fit of the Planck mission data \cite{Abbott:2019yzh}.
The results of the fits are given in Table \ref{tab: fit m + cov + inhP + + inhR + homP + homR + 0.15}, and in  in Fig.(\ref{fig: Pantheon mVsz + void + homo + Planck + Riess, no shells, zsup=0.15}). 

The model $m^{\textup{Inh}}(H_0^P)$ provides the best fit of the low redshift SNe data, while $m^{\textup{Inh}}(H_0^R)$ is disfavored, in agreement with previous studies \cite{Kenworthy:2019qwq}.
The density contrast of the best fit under-density is not large, and the hedge is located around the depth of the catalog used in \cite{Riess:2016jrr} for the RC. This is supporting the argument \cite{Romano:2016utn} that the apparent tension between $H_0^P$ and $H_0^R$ could be the consequence of a local inhomogeneity whose effects have not been removed by RC, because its size is  comparable to the 2M++ depth.
As noted in \cite{Romano:2016utn}, the  high redshift luminosity distance is not affected by the local inhomogeneity, since its effect is proportional to the volume average of the density contrast, which becomes negligible at high redshift.
The inhomogeneity parameters are in  good agreement with recent  results  of number counts analysis \cite{wong2021local}, further supporting its existence.

This kind of under-density could have been seeded by a peak of primordial curvature perturbations \cite{Romano:2014iea}. 
The statistical evidence of its presence should be considered independently of the theoretical prediction of the  probability of its occurrence \cite{Odderskov:2014hqa,Ben-Dayan:2014swa,Marra:2013rba}, i.e. the existence of inhomogeneities should be investigated using observational data rather than being excluded a priori from the analysis, on the basis of theoretical predictions. 

The value of $M$ is the key element in detecting or not the presence of the inhomogeneity in the SNe data. The value of $M^R$ is obtained assuming homogeneity and could be underestimated due to the unaccounted effects of a local under-density, as shown in Eq.(\ref{DM}), causing the well known Hubble tension.

\begin{figure}[H]
    \centering
    \includegraphics[scale=0.43]{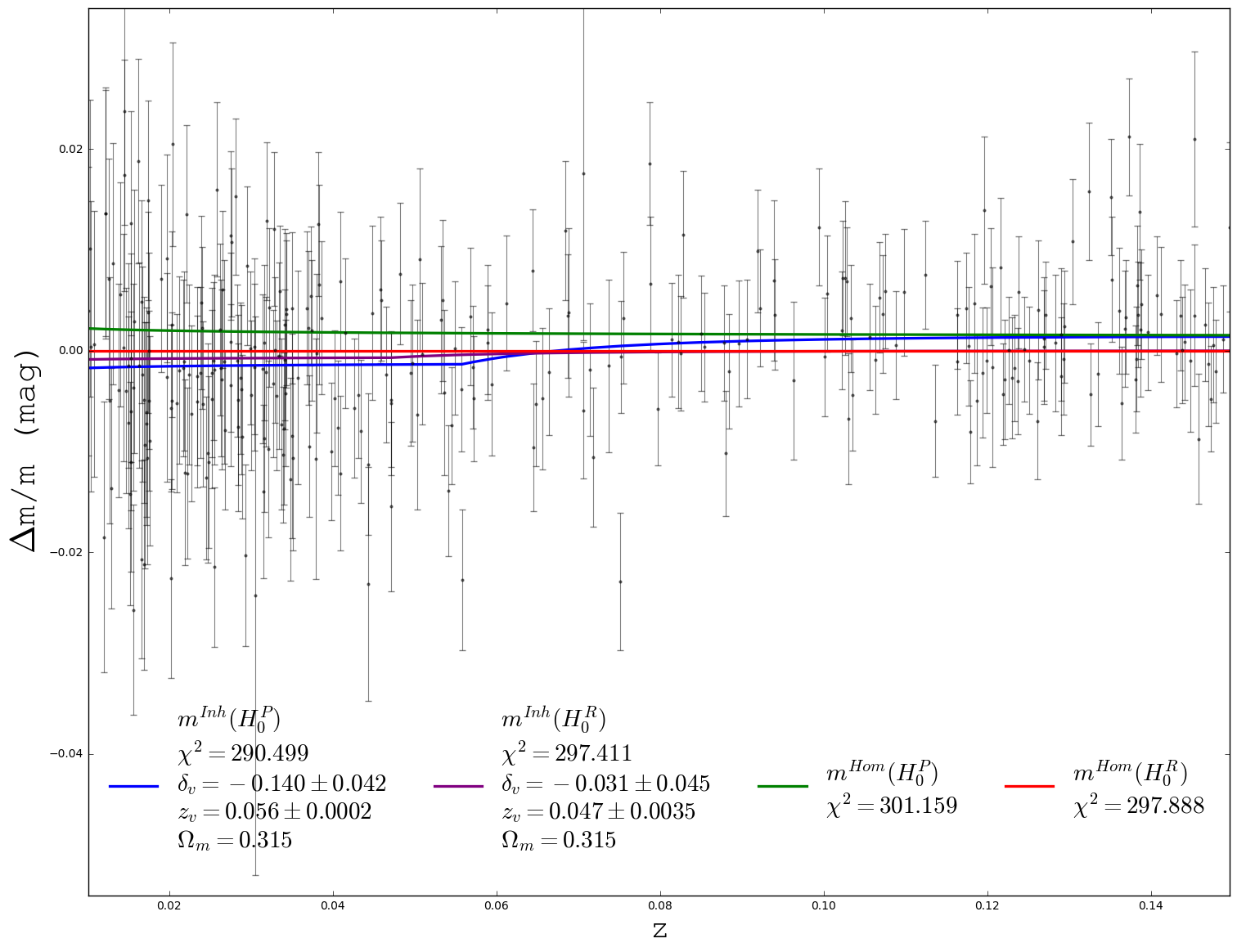}
    \caption{Fit of the Pantheon dataset with $z<z_{sup}\equiv 0.15$. The relative error of $m$ is defined as  $\Delta m/m=(m-m^{\textup{Hom}}(H_0^R))/m^{\textup{Hom}}(H_0^R)$ . The best fit is $m^{\textup{Inh}}(H_0^P)$ and the model $m^{\textup{Hom}}(H_0^P)$ does not provide a good fit of the data, in agreement with \cite{Kenworthy:2019qwq}, based on the parameters $\{H_0^P,M^P\}$, is the best fit model.}
    \label{fig: Pantheon mVsz + void + homo + Planck + Riess, no shells, zsup=0.15}
\end{figure}

\begin{table}[H]
\label{covM}
\centering

\begin{tabular}{lccc c ccc c c}
\hline
 & \multicolumn{4}{c}{$z_{sup}=0.15$} &&  \\
\cline{2-9} 
& $\delta_v$   & $z_{v}$  & $\chi^2$ & $\chi^2_{red}$ & AIC & $\Delta $ AIC & BIC& $\Delta $ BIC \\
\cline{2-9} 

$m^{\textup{Inh}}(H_0^P)$ & $-0.140 \pm 0.042 $ & $0.056 \pm 0.0002$  & 290.499 & 0.985 & 294.499 &    3.4 & 304.4082 & 7.3889 \\
$m^{\textup{Inh}}(H_0^R)$ & $-0.031 \pm 0.045 $ & $0.047 \pm 0.0035$  & 297.411 & 1.008 & 301.411 & -3.5 &311.3202 & 0.4769\\
$m^{\textup{Hom}}(H_0^P)$ & -                   & -                   & 301.159 & 1.014 & 301.159 & -3.3& 315.0682 & -3.2710 \\
$m^{\textup{Hom}}(H_0^R)$ & -                   & -                   & 297.888 & 1.003 & 297.888 & 0& 311.7972 & 0 \\

\hline
\end{tabular}
               
    \caption{Results of the fit of the Pantheon data.  The inhomogeneous models have two extra parameters, $\{z_v,\delta_v\}$, with respect to the homogeneous models, which are taken into account in the calculation of $\chi^2_{red}=\chi^2/d.o.f.$ and $AIC$. The $\Delta AIC$ and $\Delta BIC$ are defined with respect to the homogeneous model $m^{\textup{Hom}}(H_0^R)$. The inhomogeneous model $m^{\textup{Inh}}(H_0^P)$ is the best fit model, according to  the $\chi^2_{red}$, the $AIC$, and the $BIC$, while $m^{\textup{Inh}}(H_0^R)$ is disfavored, in agreement with \cite{Kenworthy:2019qwq}. This shows that a small local inhomogeneity allows to fit well SNe calibrated with the value of the absolute magnitude $M^P$ implied by  $H_0^P$. Analyzing low redshift observations ignoring the presence of such an inhomogeneity  can lead to a miss-estimation of $M$, and the consequent apparent Hubble tension. The latter is in fact due to the $M$ tension, i.e. the  difference between $M^R$ and $M^P$. If the effects of the inhomogeneity were taken into account, the $M$ tension, and consequently the $H_0$ tension, could be removed. }
    \label{tab: fit m + cov + inhP + + inhR + homP + homR + 0.15}
\end{table}

\section{Conclusions}

We have analyzed low redshift SNe data with different cosmological models.
We have found that a model with a small local under-density with $H_0=H_0^P$ can fit the data better than a homogeneous model with $H_0=H_0^R$.
The parameters of the inhomogeneity we have obtained are in good agreement with number counts observations \cite{wong2021local}.

The existence of this local under-density, if not taken  into proper account, can produce a miss-estimation of all background cosmological parameters obtained under the assumption of large scale homogeneity, and it can explain for example the Hubble tension \cite{Romano:2016utn}. 
This is in agreement with the theoretical prediction of a local inhomogeneity effects, whose leading monopole perturbative contribution is proportional to the volume averaged density contrast, implying that the high redshift luminosity distance is not affected, including the distance of the last scattering surface from which the $H_0$ is estimated with CMB observations.
It is remarkable that the edge of the inhomogeneity is located around the depth of the 2M++ catalog, used to compute the peculiar velocity redshift correction.  This naturally explains why the redshift correction applied to Pantheon data is not able to  remove the effects of the inhomogeneity obtained in our analysis.

The analysis presented in this paper is not fully Bayesian, since the values of $M$ are fixed without considering the effects of their respective errors. While this approach can work as a first approximation, it would be important to confirm the results with a full Bayesian analysis. 
In the future it will also be interesting to fit independently the values of $H_0$ and $M$ using low red-shift observations, without using any prior.
It will also be important to confirm our results with a joint fit of other observables such as as number counts\cite{Keenan:2013mfa,wong2021local},  to include the effects of possible anisotropies, and to analyze data of higher redshift SNe.

\appendix

\section{Model independent consistency relation between $H_0$ and $M$}
From the relation between the distance modulus and the luminosity distance
\bea
\log_{10}(D_L) &=& 1 + \frac{\mu}{5} \,, \\
\mu&=&m-M \,, \label{muD}
\eea
we obtain  
\bea
m^{obs}&=&(5 \log{d}-5)+(M-5\log{H_0})=l(z)+g(M,H_0)\,, \\
l(z)&=&5 \log{d(z)}-5 \,,  \\
g(M,H_0)&=&M-5\log{H_0} \label{fg} \,,
\eea
where we have defined $D_L=d/H_0$, and the function $l(z)$ could be an arbitrary function of the redshift, not necessarily that corresponding to a $\Lambda\rm CDM$ model.
For a flat $\Lambda\rm CDM$ Universe we have for example
\bea
h(z)&=& \left[\Omega_m(1+z)^3+\Omega_{\lambda}\right]^{1/2} \,,\\
d(z)&=& (1+z)\int^z \frac{dz'}{h(z')} \,,\\
D_L(z)&=&\frac{1}{H_0}d(z) \,.
\eea

It is evident from Eq.(\ref{fg})  that there can be a degeneracy between the parameters $H_0$ and $M$, since for the same $l(z)$, different combinations of $\{H_0,M\}$ can give the same $m^{obs}$, as long as $g=const$.
The parameters $\{\Omega_i,H_0,M\}$ are in general independent, and a joint analysis is required to obtain the best fit values.
For $\Lambda\rm CDM$ models the function $l(z)$ is only mildly dependent on  $\Omega_i$ at low-redshift because 
\bea
D_L(z)&=&\frac{1}{H_0}\left(z+\frac{1-q_0}{2}z^2+ ..\right) \,, \\
q_0&=&\frac{3}{2}\Omega_m-1 \,,
\eea
implying that $d\approx z$, which is approximately independent of $\Omega_i$, i.e. we get the Hubble's law. For this reason only high redshift observations can provide evidence of dark energy, because only higher order terms in the Taylor expansion of $D_L(z)$  depend on $\Omega_i$. 

Let's consider two models with the same function $d$, $d_a=d_b$, where we are denoting with subscripts $a,b$ quantities corresponding to the two models. For example these could be $\Lambda\rm CDM$ models with the same parameters $\Omega_i$.

Under the  assumption  $d_a=d_b$  we get 
\be
g_a=g_b=M_a-5\log_{10}{H_a}=M_b-5\log_{10}{H_b}  \,,
\ee
from which 
\bea
    D_L^{a} &=& D_L^{b} \frac{H_a}{H_b} \,, \\ \label{HM}
    \mu_{a} &=& \mu_{b} + 5\log_{10}\left(\frac{H_{b}}{H_{a}}\right) \,, \\
    M_{a} &=& M_{b} + 5\log_{10}\left(\frac{H_{a}}{H_{b}}\right) \,, \label{MH0}
\eea
i.e. $\{D_L^{a},H_{a},M_a\}$ and $\{D_L^{b},H_{b},M_b\}$. The last equation gives a \emph{consistency relation} between the values of $H_0$ and $M$ for different models. This relation is model independent because it only assumes Eq.(\ref{muD}).

The derivation of the consistency relation given in Eq.(\ref{MH0}) is also based  on  assuming that $\{d_a=d_b \, \rightarrow \, g_a=g_b\}$. In this paper we apply the formulae to $\Lambda\rm CDM$ models with the same $\Omega_i$, so the assumption $d_a=d_b$ is exact at any redshift, but even if the $\Omega_i$ were different, at low redshift it could still be safely applied as explained above, because the relation $d\approx z$ is in good agreement with observations in that range,  independently of the values of the parameters $\Omega_i$. The same would apply to any other  cosmological model in agreement with observations, not necessarily a $\Lambda\rm CDM$ model. For example in the case of a FRW model compared to a FRW+void model, fitting the same low redshift observations, i.e. with $m_a=m_b$, and $d_a=d_b\approx z$, where $H_a=H_b$ are the local slopes of $D_a=z/H_a$ and $D_b=z/H_b$. 
The effects of the inhomogeneity on the luminosity distance lead to a correction to the Hubble parameter approximately given by  \cite{Romano:2016utn}
\be
\frac{\Delta H_0}{H_0}=-\frac{1}{3}f \overline{\delta}(z) \,,
\ee
where $\overline{\delta}$ is the volume average of the density contrast. The above formula shows that an under-density increases the local estimation of $H_0$ with respect to the background value.

Considering a set of low redshift SNe we can also derive the effect of a local inhomogeneity on the absolute magnitude $M$
\bea
\Delta M &=& 5\log_{10}\frac{H_0+\Delta H_0}{H_0} =5\log_{10} \left(1-\frac{1}{3}f \overline{\delta}(z)\right) \,.
\eea

An under-density is expected to induce a positive correction to $M$, in agreement with Table \ref{tab:pantheon_cal}. If a local under-density is present, and redshift correction cannot completely remove its effects on the distance of the anchors, the value of $M^{Hom}=M^{true}+\Delta M$, obtained analyzing data under the  assumption of homogeneity, would be larger than the true value  $M^{true}$. Using this value as prior for $M$, or using as prior the value of $H_0$ obtained from it \cite{Riess:2016jrr}, leads to an  apparent $H_0$ tension, which is in fact the consequence of a $M$ overestimation.

For the Pantheon data set the parameters $\{H_0= 73.24 \pm 1.59,M=-19.25 \pm 0.71\}$ from \cite{Riess:2016jrr} have been taken as reference.
The values for the parameters obtained using the formulae above are given in Table  \ref{tab:pantheon_cal}, including the magnitude $M^P$ corresponding to $H_0^P$. 
\begin{table}[H]
    \centering
   \begin{tabular}{lcc}
   \hline
\toprule
Dataset &$ H_0 (km \, s^{-1} Mpc^{-1})$ & M  \\
\midrule
\hline
Riess & \underline{ $73.24 \pm 1.59$ } & \underline{$-19.25 \pm 0.71$}\\
Planck & \underline{ $67.4 \pm 0.5$} & $-19.4 \pm 0.65$  \\
\bottomrule
\hline
\label{Tpar}
\end{tabular}
    \caption{Values of $\{H_0,M\}$ used in analysis SNe data, derived using the values in \cite{Riess:2016jrr} as reference. The first row shows the values from  \cite{Riess:2016jrr}, and the second row the value of $H_0$ from \cite{Akrami:2018odb} and the implied value of $M$ obtained using Eq.(\ref{HM}). The  publicly available values are underlined, while the value of $M$ for Planck, inferred using Eq.(\ref{HM}), is not underlined.}
    \label{tab:pantheon_cal}
\end{table}

This procedure is not always correctly performed in the literature, producing to an \emph{implicit bias in selecting models}.

\section{Anchors distance and absolute magnitude miss-estimation }

Since we can only measure the apparent magnitude of SNe, a calibration is needed to estimate the absolute luminosity $M$.
For low redshift SNe the distane is obtained from the period-luminosity relation for Cepheids in the same galaxy of the SNe, which is calibrated using the angular diameter distance as anchor, the NGC4258 megamaser  \cite{Reid:2019tiq}. If the local Universe were not homogeneous the NGC4258 angular diameter distance would also be affected, implying a miss-estimation of its distance, and consequently of $M$.

As shown in \cite{Romano:2016utn}, the effects of a local inhomogeneity on the high redshift luminosity distance are negligible, since they are proportional to the volume average of the density contrast, making the effects only important for low redshift observations.

As shown in appendix A, the absolute magnitude can be miss-estimated due to the unaccounted effects of a local inhomogeneity, according to
\bea
\Delta M &=& 5\log_{10} \left(1-\frac{1}{3}f \overline{\delta}(z)\right) \,,
\eea
showing that an under-density can lead to an overestimation of the anchors distance, and consequently to an overestimation of the value of $M$ obtained using those anchors.
The procedure to estimate $M$ \cite{Riess:2016jrr} is in fact assuming that all inhomogeneities effects have been removed by applying the RC, but if that is not the case, then also the value of $M$ would receive a correction, not just $H_0$.

\section{Volume average $\overline{\delta}(z)$ of a step density contrast} 

For a step density contrast profile of the type given in Eq.(\ref{deltastep}), using the low redshift approximation $\chi\approx z/(a H_0)=z(1+z)/H_0$,  the volume average of the density contrast over a sphere of comoving radius $\chi(z)$ is
\be
\overline{\delta}(\chi)=\frac{3}{4 \pi \chi^3}\int^{\chi}_0 4 \pi \chi'^2 \delta(\chi') d\chi' =\frac{ H_0^3}{[z(1+z)]^3} \delta_{v}\chi_v^3= \delta_{v} \left[ \frac{z_v(1+z_v)}{z(1+z)} \right]^3  \,.
\ee
The factor $(1+z_v)/(1+z)$ can be neglected in the low redshift regime, as shown in Fig.(\ref{deltaz}).




\bibliographystyle{h-physrev4}
\bibliography{Bibliography} 

\end{document}